\documentclass[%
 reprint,
 amsmath,amssymb,
 aps,
]{revtex4-1}

\usepackage[dvipdfmx]{graphicx}
\usepackage{here}
\usepackage{dcolumn}
\usepackage{bm}

\newcommand{\feses}{FeSe$_{1-x}$S$_x$}
\newcommand{\tc}{$T_{\rm c}$}
\newcommand{\TT}{$1/T_{1}T$}
\newcommand{\se}{$^{77}$Se}

\begin{document}

\preprint{APS/123-QED}

\title{Two different superconducting states and
possible antiferromagnetic quantum critical points in S-doped FeSe under pressure}

\author{T. Kuwayama$^1$, K. Matsuura$^2$, Y. Mizukami$^2$, S. Kasahara$^3$, Y. Matsuda$^3$, T. Shibauchi$^2$, Y. Uwatoko$^4$ and N. Fujiwara$^1$}
\email{Email:naoki@fujiwara.h.kyoto-u.ac.jp}
\affiliation{$^1$Graduate School of Human and Environmental Studies, Kyoto University, Yoshida-Nihonmatsu-cyo, Sakyo-ku, Kyoto 606-8501, Japan\\
$^2$Graduate School of Frontier Sciences, University of Tokyo, 5-1-5 Kashiwanoha, Kashiwa, Chiba 277-8581, Japan\\
$^3$Division of Physics and Astronomy, Graduate School of Science, Kyoto University, Kitashirakawa Oiwake-cho, Sakyo-ku, Kyoto 606-8502, Japan\\
$^4$Institute for Solid State Physics, University of Tokyo, 5-1-5 Kashiwanoha, Kashiwa, Chiba 277-8581, Japan}

\date{\today}
\begin{abstract}
We performed \se-NMR measurements on \feses \, $(x = 0.12)$ up to 3.0 GPa at an applied magnetic field of 6.02 T, and found that the superconducting (SC) phase exhibits a remarkable double-dome structure in the pressure($P$)-temperature($T$) phase diagram which is hidden at 0 T.
From the relaxation rate $1/T_1$ divided by $T$, \TT, a Lifshitz transition may occur at 1.0 GPa, and the dominant nesting vector could change due to topological changes in Fermi surfaces.
In other words, two types of antiferromagnetic (AFM) fluctuations would exist in the $P$-$T$ phase diagram.
We conclude that the SC double domes in 12\%-S doped FeSe consist of two SC states each of which correlates to a different type of  AFM fluctuation.
Furthermore, the strong AFM fluctuation at ambient pressure could originate from a possible hidden AFM quantum critical point.

\end{abstract}
\pacs{Valid PACS appear here}
\maketitle
Recently, iron chalcogenides, so-called 11 systems, have received much attention because of their unique phase diagrams.
In particular, FeSe undergoes nematic and superconducting (SC) transitions at 90 and 9 K, respectively, without any magnetism at ambient pressure \cite{Baek2015}, while an antiferromagnetic (AFM) phase exists in most iron based superconductors, such as undoped or low carrier doped 1111 and 122 systems \cite{Fernandes2014}.
The pressure $(P)$- temperature $(T)$ phase diagram for FeSe has been obtained from the resistivity measurements \cite{Sun2016a}: the nematic phase disappears at 1.5 GPa, and an AFM phase with a dome structure is induced in the $P$-$T$ phase diagram instead.
The AFM phase overlaps the nematic phase at the boundary in the $P$-$T$ phase diagram.
The SC phase develops remarkably as pressure increases above 1.5 GPa: SC transition temperature (\tc ) of 9 K at ambient pressure goes up to 37 K at 6.0 GPa.
In this pressure-induced AFM phase, a stripe-type spin configuration with the nesting vector ($\pi$, 0) has been suggested from NMR measurements \cite{Wang2016}.
A theoretical investigation proposed that an inner hole-like pocket appears due to increasing pressure and it would make the AFM ordering with the $(\pi, 0)$ nesting \cite{Yamakawa2017a}.
More information about the Fermi surfaces at ambient pressure has been obtained from the angle resolved photo emission spectroscopy (ARPES) \cite{Shimojima2014, Kasahara2014, Suzuki2015, Watson2015a, Watson2017, Kushnirenko2018, Coldea2018}.
The Fermi surfaces of the pure FeSe are constructed by a hole-like pocket at the $\Gamma$ point and elliptical electron-like pockets at the M point.
Several experiments suggest orbital ordering under the nematic states, where the degeneracy between $d_{xz}$ and $d_{yz}$ orbitals is resolved \cite{Shimojima2014, Watson2015a}.

\begin{figure*}[tb]
	   \centering
	   \includegraphics[width=16cm]{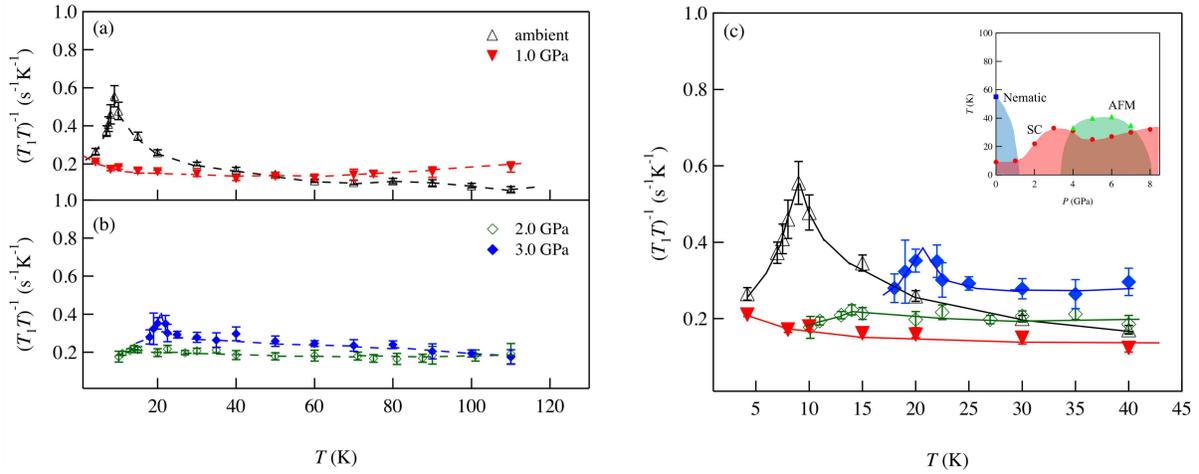}
	   \caption{
	   The relaxation rate $1/T_1$ divided by temperature $(T)$, \TT , for \se \, measured at 6.02 T up to 3.0 GPa.
	   The dashed and solid lines are the guides for the eyes.
	   (a), (b) \TT \, at low and high pressures, respectively.
	   (c) Expansion of \TT \, for Figs.\ref{figure:T1T}a and \ref{figure:T1T}b at low $T$ below 60 K.
	   The inset shows the phase diagram obtained from the resistivity measurements \cite{Matsuura2017}.}
	   \label{figure:T1T}
	\end{figure*}

The phase diagram determined from the resistivity dramatically changes with sulfur (S) doping \cite{Matsuura2017}: the pressure-induced AFM phase with the dome structure moves to a higher pressure region as the doping level is increased.
As a result, the nematic phase is segregated from the AFM phase in the $P$-$T$ phase diagram.
Interestingly, \tc \, for $x = 0.12$ reaches a maximum ($\sim$30 K) at the intermediate pressure ($\sim$3 GPa) where both the nematic and AFM phases are absent (see the inset of Fig.\ref{figure:T1T}c).
Contrary to the $P$-$T$ phase diagram, no AFM phases are induced in the $x$-$T$ phase diagram at ambient pressure \cite{Watson2015b, Hosoi2016, Reiss2017}.
An additional hole-like pocket emerges, and the electron-like pockets become isotropic as the doping level is increased \cite{Coldea2016, Watson2015b, Reiss2017, Skornyakov2018}.
Because the nematic, SC, and AFM phases overlap each other in a complex manner in the $P$-$T$ phase diagram for the pure sample, the 12\%-S doped sample is preferred for the investigation of the origin of a high \tc \, under pressure.

In the present work, we revealed the double-dome structure of the SC phase which is hidden at 0 T from \se \,-NMR measurements on 12\%-S doped FeSe under an applied field.
We found that each of these two domes correlates to a different type of AFM fluctuation.
In addition, we argue that the strong AFM fluctuation observed at ambient pressure could originate from a possible hidden AFM quantum critical point (QCP).

	\begin{table}[b]
	\caption{\label{table1} \tc \, measured at 0 T and 6.02 T up to 3.0 GPa.}
	\begin{ruledtabular}
	\begin{tabular}{ccc}
	 Pressure (GPa) & \tc \, at 0 T (K)& \tc \, at 6.02 T (K)\\
	\hline
	ambient & 9.8 & 6.5 \cite{Xiang2017} \\
	1.0 & 8.8 & 2 \cite{Xiang2017}\\
	2.0 & 15.5 & 14.2 \\
	3.0 & 25.5 & 24.1 \\
	\end{tabular}
	\end{ruledtabular}
	\end{table}
We performed \se-NMR measurements at 6.02 T up to 3.0 GPa on a 12\%-S doped single crystal with dimensions of about $1.0 \times 1.0 \times 0.5$ mm.
We used a NiCrAl pressure cell \cite{Fujiwara2007} and Daphne oil as pressure mediation liquid.
We placed the crystal in the pressure cell so that the FeSe plane was parallel to the applied field.

First, we determined \tc \, from AC susceptibility measurements using the tank circuit of a NMR probe at both 0 T and 6.02 T (see Table \ref{table1}).
In the absence of a field, \tc \, is enhanced with increasing pressure and reaches about 25 K at 3.0 GPa.
Surprisingly, at 1.0 GPa, no SC phase was observed down to 4.2 K at 6.02 T.
This remarkable suppression of \tc \, at 1.0 GPa is also observed in the NMR measurements as mentioned below.

We measured the relaxation time $T_1$ using a conventional saturation recovery method.
The relaxation rate provides a measure of low-energy spin fluctuations.
When the wave vector $(q)$ dependence of the hyperfine interaction is neglected, \TT \, is expressed as follows:
	\begin{equation}
	\frac{1}{T_1T} \sim \sum_q \frac{{\rm Im} \chi (q, \omega)}{\omega}
	\end{equation}
where $\omega$ and $\chi (q, \omega)$ represent the NMR frequency and the dynamical spin susceptibility, respectively.
Fig.\ref{figure:T1T}a and Fig.\ref{figure:T1T}b show \TT \, in low and high pressure regions, respectively.
The temperature where \TT \, has a peak is in good agreement with \tc \, determined from the AC susceptibility measurements.
There is no peak at 1.0 GPa down to 4.2 K, which is also consistent with the AC susceptibility measurements at 6.02 T.
Below \tc , the signal intensity becomes extremely small, and we could not detect signals below 10 K at 2.0 GPa and 15 K at 3.0 GPa.
At ambient pressure, \TT \, clearly shows Curie-Weiss-like behavior below 60 K where the system undergoes the nematic transition.
Thus, the strong AFM fluctuations exist at ambient pressure.
At 1.0 GPa, the strong AFM fluctuations observed at ambient pressure are strongly suppressed (Fig.\ref{figure:T1T}a).
From 2.0 GPa to 3.0 GPa, \TT \, is enhanced with increasing pressure, and at 3.0 GPa, the Curie-Weiss-like behavior revives below 30 K (Fig.\ref{figure:T1T}b).
Because the system comes close to the magnetic ordering state with increasing pressure (see the inset of Fig.\ref{figure:T1T}c), one would expect that \TT \, increases monotonically due to the AFM fluctuations.
However, \TT \, is reduced up to a pressure of 1.0 GPa and then recovers as further pressure is applied.
This unexpected $P$ dependence of \TT \, can be clearly seen in Fig.\ref{figure:T1T}c.

\begin{figure}[tbp]
	      \centering
	   \includegraphics[width=8.5cm]{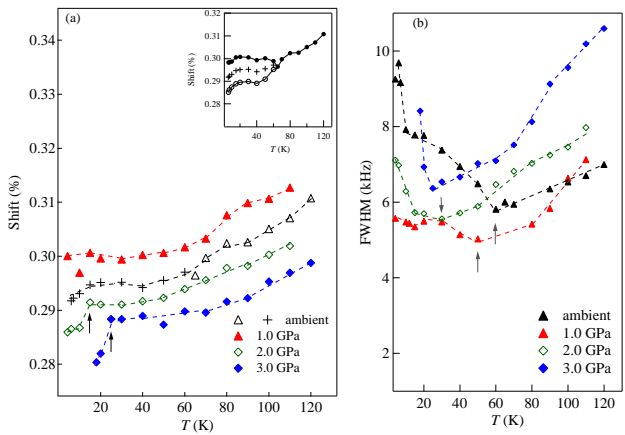}
	   \caption{(a) $T$ dependence of the \se-NMR shift at several pressures.
	   The black crosses represent the average of two lines in the nematic phase.
	   The black arrows represent \tc.
	   The inset shows the $T$ dependence of the shift at ambient pressure.
	   The split lines are obtained from two Gaussian fits.
	   (b) $T$ dependence of the full width at half maximum (FWHM) determined from a single Gaussian fit.
	   The gray arrows show the nematic transition points determined from onsets of the upturn.}
	   \label{FWHM}
	\end{figure}

A single \se-NMR signal was observed in a tetragonal state, and it exhibits a double-peak structure in the nematic phase.
At ambient pressure, a \se \, line splits into two lines below 60 K, which is in good agreement with the nematic transition temperature obtained from the resistivity measurements \cite{Matsuura2017, Xiang2017}.
For the pure sample $(x = 0)$, the signal was detected as two separated lines in the nematic phase \cite{Baek2015, Wang2016, Wiecki2018}.
However, for our 12\%-S doped sample, the two lines overlap each other.
This implies that the Fermi surfaces become isotropic due to S doping.
\se-NMR shifts at ambient pressure determined from FFT-NMR spectra are shown in the inset of Fig.\ref{FWHM}a.
The average of the shifts at ambient pressure are plotted as black crosses in the main panel of Fig.\ref{FWHM}a.
The shifts in Fig.\ref{FWHM}a qualitatively exhibit similar $T$ dependence, and the quantitative difference comes from the density of states.
Thus, the Fermi surfaces would change by the application of pressure.
The NMR shift undergoes a sudden drop at about \tc \, at high pressures of 2-3 GPa, which implies that the SC gap structure changes between ambient pressure and 2-3 GPa.

\begin{figure}[tbp]
	      \centering
	   \includegraphics[width=8cm]{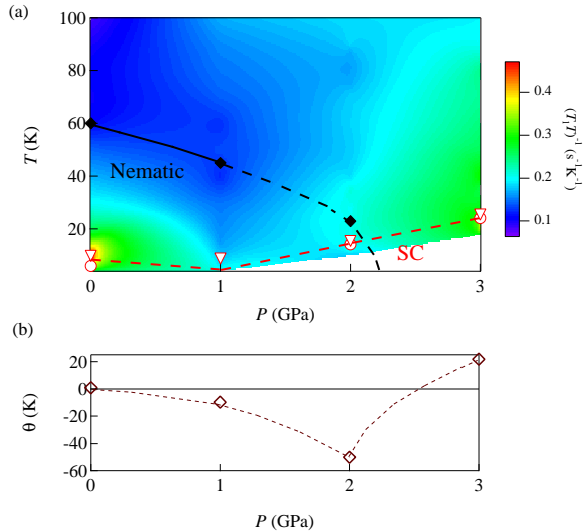}
	   \caption{(a) Phase diagram and magnetic fluctuations in \feses \, $(x = 0.12)$.
	   The inverted triangle and circles represent \tc \, at 0 T and 6.02 T, respectively.
	   The value of \TT \, is superimposed in the phase diagram by a colored contour.
	   The black squares represent the long and short range nematic transition temperatures determined from the FWHM.
	   (b) The Weiss temperature $\theta$ determined from the Curie-Weiss fit for \TT.
	   The dashed line is the guide for the eyes.}
	   \label{weiss}
	\end{figure}

Fig.\ref{FWHM}b shows the full width at half maximum (FWHM) which is obtained from a single Gaussian fit for the \se \, line.
Because the hydrostaticity deteriorates, the FWHM tends to increase with increasing pressure.
At ambient pressure, where the system undergoes the nematic phase below 60 K, the FWHM has two inflection points.
The first point at about 60 K reflects the nematic transition and the second point at about 9 K reflects the SC transition.
In addition, the $T$ dependence of the FWHM at ambient pressure shows a convex upward characteristic from 60 to 9 K, and a similar $T$ dependence is observed at 1.0 GPa from 50 to 15 K.
At 2.0 GPa, the FWHM takes a slight upturn at about 30 K which is represented by the gray arrow in Fig.\ref{FWHM}b.
From the resistivity measurements \cite{Matsuura2017, Xiang2017}, the nematic QCP may exist near 1.0 GPa.
Thus, the upturn seen in the FWHM in a pressure of 1-2 GPa may reflect short range nematic order suggested in pure FeSe \cite{Wiecki2017a, Wang2017c}.
At 3.0 GPa, the $T$ dependence of the FWHM shows a single upturn at \tc, so that the nematic phase is completely absent.
The long or short range nematic transition temperature is plotted in Fig.\ref{weiss}a.


As we will discuss below, our results imply that (1) two distinct SC states in \feses \, $(x = 0.12)$ form the double-dome structure in the $P$-$T$ phase diagram under the applied magnetic field and that (2) a hidden AFM QCP exists at ambient pressure.

First, we discuss the former, (1).
At first glance, \tc \, seems to be suppressed by the nematic fluctuations around a nematic QCP and seems to increase at higher pressures because the SC state is free from the suppression.
From the FWHM (Fig.\ref{FWHM}b), however, the nematic fluctuation would exist up to 2.0 GPa whereas an anomaly of \tc \, emerges at 1.0 GPa (Fig.\ref{weiss}a).
It is possible that a Lifshitz transition causes the anomaly in \tc \, at 1.0 GPa.
Considering the $P$ and $T$ dependence of \TT \, (Fig.\ref{weiss}a), AFM fluctuations seem to correlate to \tc.
To understand the relationship between the AFM fluctuations and the SC phase, the evolution of the Fermi surfaces and the nesting vector upon S doping and the application of pressure should be taken into account.
From the ARPES and quantum oscillation measurements, the electron-like pockets at M point become isotropic with S doping and a Lifshitz transition may occur \cite{Reiss2017, Phan2017, Coldea2016}.
Furthermore, the theoretical investigation suggests that the additional hole-like pocket emerges at the $\Gamma$ point under pressure \cite{Yamakawa2017a, Skornyakov2018}.
Both S doping and the application of pressure would make the electron pockets isotropic \cite{Skornyakov2018, Coldea2018}.
Taking these topological changes into account, in S doped FeSe under pressure, remarkable changes of the Fermi surfaces and the dominant nesting vector could occur.
The NMR spectra reflect these topological changes as mentioned.
Thus, two types of AFM fluctuations (colored in green in Fig.\ref{weiss}a) imply that the Lifshitz transition occurs and the dominant nesting vector changes around 1.0 GPa.
The $P$ dependence of \tc \, correlates to this change in the topology, which can be clearly seen in Fig.\ref{weiss}a.

The \se-NMR shifts are also consistent with the scenario that two different SC states form the double-dome-like SC phase in the $P$-$T$ phase diagram at 6.02 T.
In the SC phase, the shift decreases with decreasing $T$.
Contrary to the shift at ambient pressure, it exhibits a remarkable drop at \tc \, at 2.0 GPa and 3.0 GPa.
While the data points  to detect the SC gap symmetry are few, this difference may reflect the two different SC phases.
From the above, two different SC phases exist in \feses \, $(x=0.12)$ under pressure, and thus, SC-SC transition may occur.
Recently, the STM and STS measurements imply two distinct pairing states in the $x$-$T$ phase diagram \cite{Hanaguri2017, Sato2017}.
It is uncertain that the SC-SC transitions in the $P$-$T$ and $x$-$T$ diagrams are the same at present.

	\begin{figure}[tbp]
	      \centering
	   \includegraphics[width=6cm]{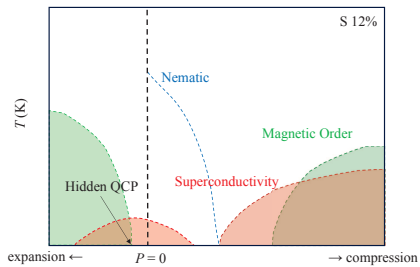}
	   \caption{Schematic phase diagram for \feses \, $(x = 0.12)$.
	   The black dashed vertical line represents ambient pressure.}
	   \label{phase}
	\end{figure}
Next, we discuss a possible hidden AFM QCP, (2).
In FeSe systems, the AFM fluctuation is very strong at ambient pressure despite the absence of magnetic order, which is an open problem at present \cite{Imai2009, Wang2016}.
Herein, we propose a possible explanation for this phenomenon.
Fig.\ref{weiss}b shows the $P$ dependence of the Weiss temperature $\theta$ determined from a Curie-Weiss fit:
	\begin{equation}
	 \frac{1}{T_1T} \sim a + \frac{b}{T-\theta}
	\end{equation}
where $a$ and $b$ are assumed to be coefficients independent of $T$.
In general, $\theta = 0$ K at QCP.
In Fig.\ref{weiss}, $\theta$ takes 0 K near 0 GPa and 2.6 GPa.
The latter is accountable as the QCP of the pressure-induced AFM phase, although the value is smaller than that suggested from the resistivity measurements \cite{Matsuura2017}.
It is important to ascertain why $\theta$ takes 0 K near 0 GPa where no magnetism is observed.
A possible explanation is that another AFM phase would exist in an imaginary negative pressure region.
Application of hydrostatic negative pressure is unrealistic.
In terms of lattice expansion, however, isovalent substitutions can give a clew for understanding the unexpected quantum criticality at 0 GPa.
For example, Te-doped FeSe and FeTe have larger lattice constants than S-doped FeSe and pure FeSe \cite{Mizuguchi2010, Paulose2010, Bao2009}.
FeTe$_{1-y}$Se$_y$ undergoes an AFM transition with small values of $y$ \cite{Mizuguchi2010, Paulose2010, Koteski2017}.
According to a recent preprint, the AFM fluctuation is reduced as the S doping level increases at ambient pressure \cite{Wiecki2018}.
This fact can be understood if the system with excess S doping level is located far from the hidden QCP, because S doped FeSe has smaller lattice constants than pure FeSe \cite{Matsuura2017, Millican2009}.
Therefore, the strong AFM fluctuations observed in pure FeSe and \feses \, could originate from a possible hidden AFM QCP.
Interestingly, the $P$ dependence of the Weiss temperature in the low pressure region seems to have some correlations with the nematic transition temperature (see Fig.\ref{weiss}).

Assuming that the hidden QCP exists, the phase diagram is reminiscent of that for LaFeAsO$_{1-x}$H$_x$, where two SC domes are sandwiched between two AFM phases \cite{Fujiwara2013, Fujiwara2017}.
Fig.\ref{phase} shows a schematic phase diagram of FeSe$_{1-x}$S$_x$.
The double dome structure of the SC phase is sandwiched between two separated AFM phases.
The hidden QCP and the unique phase diagram are first suggested from our present NMR measurements on S-doped FeSe, where the nematic phase is segregated from the AFM phase.
Because there is no way to apply hydrostatic negative pressure, this phase diagram is speculative.
To clarify this, systematic investigations with wider $x$ and $P$ ranges are needed.
However, it is certain that we obtained clues for understanding the SC pairing mechanism in iron-based superconductors.

In summary, we have demonstrated that two different SC states form the double-dome-like SC phase in \feses \, $(x = 0.12)$ under pressure.
The Lifshitz transition may occur around 1.0 GPa where \tc \, is strongly suppressed.
Thus, an unexpected $P$ and $T$ dependence of \TT \, reflects topological changes in Fermi surfaces, that is, the dominant nesting vector below 1.0 GPa is different from that above 1.0 GPa.
In addition, the Weiss temperature $\theta$ takes 0 K near 0 GPa, which implies that another AFM QCP  exists.
In nature, the SC phase in \feses \, $(x = 0.12)$ has the double-dome-like structure, and is sandwiched between two AFM phases, similar to the case of LaFeAsO$_{1-x}$H$_x$.

The NMR work is supported by JSPS KAKENHI Grant Number JP18H01181 and a grant from Mitsubishi Foundation. We thank H. Kontani for discussion.


\end{document}